\documentclass[journal]{IEEEtran}
\usepackage[utf8]{inputenc}
\usepackage{stfloats}
\usepackage{cite}
\usepackage{amsmath}
\usepackage{amsfonts}
\usepackage{dsfont}
\usepackage{graphicx}
\usepackage{changepage}
\usepackage{color}
\UseRawInputEncoding

\ifCLASSINFOpdf

\fi

\begin{document}

\title{Performance Analysis of UAV-based Mixed RF-UWOC Transmission Systems}

\author{Sai Li, Liang Yang, and Daniel Benevides da Costa
\thanks{S. Li and L. Yang are with the College of Computer Science and Electronic Engineering, Hunan University, Changsha 410082,
China, (e-mail:lisa2019@hnu.edu.cn, liangy@hnu.edu.cn).}
\thanks{D. B. da Costa is with the Department of Computer Engineering, Federal University of Cear\'{a}, Sobral 62010-560, Brazil (email: danielbcosta@ieee.org).}}
\maketitle

\begin{abstract}
In this paper, we investigate the performance of a mixed radio-frequency-underwater wireless optical communication (RF-UWOC) system where an unmanned aerial vehicle (UAV), as a low-altitude mobile aerial base station, transmits information to an autonomous underwater vehicle (AUV) through a fixed-gain amplify-and-forward (AF) or decode-and-forward (DF) relay. Our analysis accounts for the main factors that affect the system performance, such as the UAV height, air bubbles, temperature gradient, water salinity variations, and detection techniques. Employing fixed-gain AF relaying and DF relaying, we derive closed-form expressions for some key performance metrics, e.g., outage probability (OP), average bit error rate (ABER), and average channel capacity (ACC). In addition, in order to get further insights, asymptotic analyses for the OP and ABER are also carried out. Furthermore, assuming DF relaying, we derive analytical expressions for the optimal UAV altitude that minimizes the OP. Simulation results show that the UAV altitude influences the system performance and there is an optimal altitude which ensures a minimum OP. Moreover, based on the asymptotic results, it is demonstrated that the diversity order of fixed-gain AF relaying and DF relaying are respectively determined by the RF link and by the detection techniques of the UWOC link.
\end{abstract}


\section{Introduction}
\IEEEPARstart{U}{nmanned} aerial vehicles (UAVs)-based wireless communication systems have received extensive attention, especially as aerial base stations (BSs) and/or as relays to assist existing communication systems \cite{1,2,3}. The actual scenarios of UAVs application include real-time monitoring, wireless coverage, and safety supervision. Compared with the ground-based Internet of Things (IoT) communication platforms, UAV wireless communication systems have the advantages of easy deployment, low cost, and better communication channel quality brought by short-range line-of-sight (LoS) connections. In addition, unlike fixed relays, UAV-assisted relays can flexibly adjust their locations according to the changes of urban environment to improve communication reliability and achieve long-distance data transmission \cite{4,5,6,7,8}. For instance, an UAV-based mixed radio-frequency/free-space optical (RF/FSO) communication network was considered in \cite{4} with the purpose of facilitating the uplink communication among multiple users and a ground station. In \cite{5}, a multi-hop low-altitude UAV based RF/FSO communication system was proposed and analytical expressions for the outage probability (OP) and optimal altitude were derived. The secrecy performance of UAV-assisted relaying systems was investigated in \cite{6,7}. In \cite{8}, the energy harvesting technology was applied to an UAV-based communication system and the system OP was evaluated. The works in \cite{9,10} investigated the coverage quality in areas without infrastructure coverage or when the infrastructure is damaged, such as low altitude marine monitoring and emergency communications under extreme conditions, by deploying UAVs as aerial BSs. Other works focused on addressing the UAV deployment optimization and the air-to-ground (A2G) channel modeling \cite{10,11,12}. For instance, in [10] closed-form expressions for the optimal UAV altitude that minimizes the OP of an arbitrary A2G link have been obtained, while in \cite{11} the authors derived the optimal altitude that allows the UAV to reach its maximum coverage radius.

On another front, underwater wireless optical communication (UWOC) systems have widely been investigated in the literature due to their promising gains, such as wider bandwidth, higher data transmission rate, and more secure transmission \cite{13,14,15}. However, in UWOC links, in addition to absorption and scattering, underwater optical turbulence (UOT) caused by temperature fluctuations, salinity variations, and air bubbles arise as nature impediments that may degrade the system performance. To evaluate more properly their impacts on the system performance, the performance of UWOC links in the presence of different air bubble populations and/or underwater channels with temperature or salinity gradients was evaluated in \cite{14}. Specifically, in \cite{15}, by taking into account the effect of the air bubbles, temperature gradients and salinity on the water channel, the authors presented a unified UWOC turbulence model relying on the mixture Exponential-Generalized Gamma (EGG) distribution.
Recently, similar to previous RF/FSO research works, cooperative communication technology has been applied to UWOC systems in order to improve the overall reliability. In \cite{16}, employing EGG distribution, the performance of a dual-hop UWOC system was analyzed and it was shown that the proposed system is more effective and robust to underwater turbulence than a single UWOC link. The uplink scenario was considered in \cite{17}, in which the data collected by sensors located underwater was first transmitted through an UWOC link to a static relay deployed on the sea surface, and then the data was transmitted to land users through a high-speed FSO link. Assuming a downlink transmission and a terrestrial base station, the authors in \cite{18} derived closed-from expressions for the OP and average bit error rate (ABER) of mixed dual-hop RF-UWOC relaying systems. Very recently \cite{19}, the performance of RF-UWOC systems was examined assuming that the RF link and the UWOC link experience Generalized-$K$ distribution and mixture EGG distribution, respectively.

Despite the importance of all previous works, the studies related to RF-UWOC systems have mainly focused on terrestrial and underwater cooperative networks. Due to this fact, efficient coverage service of static BSs largely depends on their locations with respect to the sea surface. On the other hand, deploying UAVs as mobile aerial BSs can effectively alleviate and surpass coverage issues since UAVs are able to flexibly move to the required coverage area. Motivated by this and relying on air-to-sea cooperative communications \cite{20}, in this paper we investigate the performance of mixed RF-UWOC systems where an UAV is employed as a mobile BS and transmits information to an autonomous underwater vehicle (AUV) through a fixed-gain amplify-and-forward (AF) or decode-and-forward (DF) relay. We assume that the RF link follows Rician fading while the UWOC link undergoes mixture EGG fading distribution. To the best of the authors' knowledge, there is no work in the literature performing this kind of investigation so that our paper can serve as a benchmark for future studies. The contributions of this paper is summarized as follows:
\begin{itemize}
\item[$\bullet$] Using fixed-gain AF and DF relaying, Closed-form expressions for the cumulative distribution function (CDF) and probability density function (PDF) of the end-to-end (E2E) signal-to-noise ratio (SNR) were derived. Based on them, accurate analytical expressions for the OP, ABER, and average channel capacity (ACC) are presented. These expressions are written in terms of Extended Generalized Bivariate Fox's H-Function (EGBFHF), which can be easily implemented relying on results published elsewhere in the literature.

\item[$\bullet$] The effect of the UAV altitude and coverage on the overall system performance are examined. To this end, we derive analytical expressions for the optimal elevation angle and for the optimal UAV altitude that minimize the OP.

\item[$\bullet$]In the UWOC link, heterodyne detection (HD) and intensity modulation/direct detection (IM/DD) techniques are employed. Considering the case in which the absorption and scattering effect are not significant, as well as the UOT dominates the fading characteristics of the channel, we analyze the effects of air bubbles, temperature and salinity gradient on the performance of the proposed system setup.

\item[$\bullet$] Asymptotic expressions at high SNR regime for the OP and ABER are presented in order to get further insights. Our results reveal that the diversity order of fixed-gain AF and DF relaying are determined by the RF link and by the detection techniques of the UWOC link, respectively.
\end{itemize}

The remainder of this paper can be organized as follows. The system and channel models as well as the E2E SNRs of the fixed-gain AF and DF relaying are presented in Section II. The CDF and PDF of the two relaying strategies are derived In Section III, which will be useful for calculating the key performance metrics. Section IV derives exact expressions for the OP, ABER, and ACC along with asymptotic results. Section V provides the optimal altitude analysis for the DF relaying system. In order to obtain insightful findings, some illustrative numerical examples are conducted in Section VI. Section VII concludes the paper.

\section{System and Channel Models}
As shown in Fig. 1, we consider a dual-hop UAV-based RF-UWOC system where a UAV serves as an aerial BS to communicate with an AUV through an over-sea surface relay (R). At R, both fixed-gain AF and DF strategies are considered. In the first slot, UAV broadcasts the signal to R via the RF link (UAV-R link) and, in the second slot, the received RF signal is converted into an optical signal and then forwarded to the AUV through the UWOC link (R-AUV link). We assume that the UAV-R and R-AUV links undergo Rician fading and the mixture EGG fading, respectively.

\begin{figure}[t]
    \centering
    \includegraphics[width=3.5in]{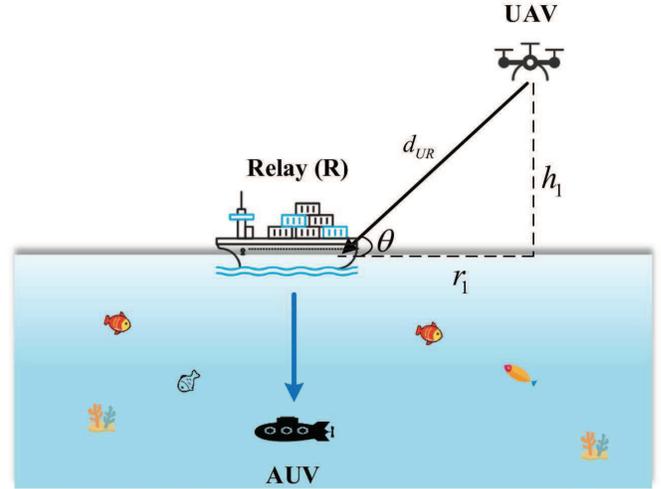}
    \caption{The dual-hop mixed UAV-based RF-UWOC system model.}
\end{figure}

\subsection{UAV-R Link}
The received signal at R is expressed as
\begin{align}
y_{UR}=\frac{1}{\sqrt{L_{UR}}}h_{UR}x+n_{1},
\tag{1}\label{1}
\end{align}
where $h_{UR}$ denotes the small-scale fading channel coefficient between the UAV and R, and $n_{1}\sim \mathcal{C}\mathcal{N}(0,N_{01})$ stands for the additive white Gaussian noise (AWGN) term with zero mean and variance $N_{01}$. Moreover, $L_{UR}=Ad_{UR}^{\alpha}$ represents the path loss, $A$ is a constant associated with the signal frequency and transmission environment, $d_{UR}=\sqrt{h_{1}^{2}+r_{1}^{2}}$ is the distance between UAV and R, where $h_{1}$ denotes the altitude of the UAV, $r_{1}$ stands for the horizontal distance between the UAV and R, $\alpha$ holds for the path loss exponent, which satisfies $\alpha(\theta)=a_{1}P_{LoS}(\theta)+b_{1}$ \cite{10}, with $\theta$ denoting the angle between the UAV and R, $P_{LoS}$ is the LoS probability, and $a_{1}$ and $b_{1}$ depend on the environment and the transmission frequency.

Therefore, the instantaneous SNR of the UAV-R link is given by $\gamma_{1}=\frac{|h_{UR}|^2}{L_{UR}}\overline\gamma_{1}$,
where $\overline\gamma_{1}$ is the average SNR of the UAV-R link. Since the UAV-R link undergoes Rician fading, the PDF of the instantaneous SNR, $\gamma_{1}$, is given by \cite{8,21}
\begin{align}
f_{\gamma_{1}}(\gamma_{1})=\frac{\vartheta e^{-K}}{\overline\gamma_{1}}\exp\left(-\frac{\vartheta \gamma_{1}}{\overline\gamma_{1}}\right)I_{0}\left(2\sqrt{\frac{K\vartheta\gamma_{1}}{\overline\gamma_{1}}}\right),
\tag{2}\label{2}
\end{align}
where $\vartheta=(1+K)L_{UR}$, $I_{0}(\cdot)$ is defined as the zero-order modified Bessel function of the first kind [22, Eq. (8.431)], $K$ means the Rician fading factor, which is modeled as a function of $\theta$. From \cite{23} and [10, Eqs. (34) and (35)], it follows that $K(\theta)=a_{2}e^{b_{2}\theta}$, where $a_{2}=K(0)$ and $b_{2}=\frac{2}{\pi}\ln\left(\frac{K(\frac{\pi}{2})}{K(0)}\right)$.
In addition, the CDF of $\gamma_{1}$ can be written as
\begin{align}
F_{\gamma_{1}}(\gamma_{1})=1-Q_{1}\left(\sqrt{2K},\sqrt{2\gamma_{1} \vartheta/ \overline\gamma_{1}}\right),
\tag{3}\label{3}
\end{align}
where $Q_{1}(\cdot,\cdot)$ is the Marcum Q-function of the first order [24, Eq. (86)].

\subsection{R-AUV Link}
\subsubsection{Amplify-and-Forward}
For the AF relaying, the received signal at AUV is expressed as
\begin{align}
y_{RA}=\eta I Gy_{UR}+n_{2},
\tag{4}\label{4}
\end{align}
where $I$ stands for the normalized irradiance, $\eta$ denotes the optical-to-electrical conversion coefficient, $G$ represents the fixed gain, and $n_{2}\sim \mathcal{C}\mathcal{N}(0,N_{02})$ denotes the AWGN term with zero mean and variance $N_{02}$.

\begin{table}[t]
\centering
\caption{Parameters of the EGG Distribution for Different Bubble Levels BL (L/min)
and Temperature Gradient ($^{\circ}\mathrm{C}.cm^{-1}$) \cite{15}}
{\begin{tabular}{p{0.4cm}p{1.2cm}p{0.5cm}p{0.5cm}p{0.5cm}p{0.5cm}p{0.5cm}p{0.7cm}}
 \hline
 \hline
BL & Temperature Gradient  &$\sigma_{I}^{2}$ & $\omega$ &$\lambda$ & a & b & c\\
 \hline
2.4 & 0.05 & 0.1484 & 0.2130 & 0.3291 & 1.4299 & 1.1817 & 17.1984 \\
2.4 & 0.20 & 0.2178 & 0.1665 & 0.1207 & 0.1559 & 1.5216 & 22.8754 \\
4.7 & 0.05 & 0.4201 & 0.4580 & 0.3449 & 1.0421 & 1.5768 & 35.9424 \\
 \hline
 \hline
\end{tabular}}
\end{table}

\begin{table}[t]
\centering
\caption{Parameters of the EGG Distribution for Different Bubble Levels BL (L/min)
for Fresh Water and Salty Water \cite{15}}
{\begin{tabular}{p{1.4cm}p{0.3cm}p{0.5cm}p{0.5cm}p{0.5cm}p{0.5cm}p{0.5cm}p{0.9cm}}
 \hline
 \hline
Salinity & BL & $\sigma_{I}^{2}$ & $\omega$ &$\lambda$ & a & b & c\\
 \hline
Salty Water & 2.4 & 0.1006 & 0.1770 & 0.4687 & 0.7736 & 1.1372 & 49.1773 \\
Salty Water & 4.7 & 0.1308 & 0.2064 & 0.3953 & 0.5307 & 1.2154 & 35.7368 \\
Salty Water & 7.1 & 0.3111 & 0.4344 & 0.4747 & 0.3935 & 1.4506 & 77.0245 \\
Salty Water & 16.5 & 1.1273 & 0.4951 & 0.1368 & 0.0161 & 3.2033 & 82.1030 \\
Fresh Water & 2.4 & 0.1088 & 0.1953 & 0.5273 & 3.7291 & 1.0721 & 30.3214 \\
Fresh Water & 4.7 & 0.1233 & 0.2190 & 0.4603 & 1.2526 & 1.1501 & 41.3258 \\
Fresh Water & 7.1 & 0.3150 & 0.3489 & 0.4771 & 0.4319 & 1.4531 & 74.3650 \\
Fresh Water & 16.5 & 1.0409 & 0.5117 & 0.1602 & 0.0075 & 2.9963 & 216.8356 \\
 \hline
 \hline
\end{tabular}}
\end{table}

\subsubsection{Decode-and-Forward}
As for the DF relaying system, the received signal at AUV can be written as
\begin{align}
y_{RA}=\eta I \hat{x}+n_{2},
\tag{5}\label{5}
\end{align}
where $\hat{x}$ denotes the decoded signal at R.

The instantaneous SNR of the R-AUV link, $\gamma_{2}$, is formulated as $\gamma_{2}=(I^r/\mathbb{E}[I]^r)\mu_{r}$, where $r$ stands for the parameter that depends on the detection technique used, i.e., $r=1$ when the HD technique is employed and $r=2$
when the IM/DD technique is adopted. The parameter $I$ is modelled by the mixture EGG distribution, whose the PDF can be given by [15, Eq. (1)]
\begin{align}
f_{I}(I){=} \frac {w}{\lambda }\exp\left({{-}\frac {I}{\lambda }}\right)
{+}(1{-}w)\frac {cI^{ac{-}1}}{b^{ac}}\frac {\exp \left({{-}\left ({\frac {I}{b}}\right)^{c}}\right)}{\Gamma (a)},
\tag{6}\label{6}
\end{align}
where $\omega$, $\lambda$, $a$, $b$, and $c$ represent the fading parameters related to the mixture EGG distribution. From (6), the $n$-th moment of $I$ can be derived as $\mathbb{E}[I^n]=\omega \lambda^{n} n!+(1-\omega)b^{n}\Gamma(a+\frac{n}{c})/\Gamma(a)$ [15, Eq. (4)]. Based on it,
the scintillation index $\sigma_{I}^{2}$ represents the turbulence intensity and can be formulated as $\sigma_{I}^{2}=\frac{\mathbb{E}[I^2]-\mathbb{E}[I]^2}{\mathbb{E}[I]^2}$. Moreover, the average electrical SNR $\mu_{r}$ is defined as $\mu_{r}=\eta^r\mathbb{E}[I]^r/N_{02}$ and the relation between $\mu_{r}$ and the average SNR $\overline\gamma_{2}$ of the R-AUV link is given by $\overline\gamma_{2}=(\mu_{r}\mathbb{E}[I^r])/\mathbb{E}[I]^r$. For the HD technique, $\mu_{1}=\overline\gamma_{2}$, while for the IM/DD technique, $\mu_{2}=\frac{\overline\gamma_{2}}{2\omega
\lambda^{2}+b^{2}(1-\omega)\Gamma(a+2/c)/\Gamma(a)}$ [15, Eq. (19)]. These parameters vary with the water temperature, water salinity, as well as bubble level. The parameters of the EGG distribution are given in Table I and Table II.

From \cite{19}, the PDF of $\gamma_{2}$ can be formulated in terms of Fox's H-function as
\begin{align}
f_{\gamma_{2}}(\gamma_{2})=\frac{\omega}{\gamma_{2}}&\, {\mathrm{H}}_{0,1}^{1,0}
\left [{{\frac{\gamma_{2}}{\lambda^r \mu_{r}}}\left |{ \begin{matrix} {-}
\\ {(1,r)} \\ \end{matrix} }\right . }\right ]\!\nonumber\\
&+\frac{(1-\omega)}{\gamma_{2}\Gamma(a)}\, {\mathrm{H}}_{0,1}^{1,0}
\left [{{\frac{\gamma_{2}}{b^r \mu_{r}}}\left |{ \begin{matrix} {-}
\\ {(a,\frac{r}{c})} \\ \end{matrix} }\right . }\right ]\!,
\tag{7}\label{7}
\end{align}
and the CDF expression can be formulated as
\begin{align}
F_{\gamma_{2}}(\gamma_{2})=\omega r&\, {\mathrm{H}}_{1,2}^{1,1}
\left [{{\frac{\gamma_{2}}{\lambda^r \mu_{r}}}\left |{ \begin{matrix} {(1,r)}
\\ {(1,r)(0,r)} \\ \end{matrix} }\right . }\right ]\!\nonumber\\
&+\frac{(1-\omega)r}{\Gamma(a)c}\, {\mathrm{H}}_{1,2}^{1,1}
\left [{{\frac{\gamma_{2}}{b^r \mu_{r}}}\left |{ \begin{matrix} {(1,\frac{r}{c})}
\\ {(a,\frac{r}{c})(0,\frac{r}{c})} \\ \end{matrix} }\right . }\right ]\!,
\tag{8}\label{8}
\end{align}
where $\mathrm{H}_{\cdot,\cdot}^{\cdot,\cdot}[\cdot]$ is the Fox's H-function [25, Eq. (1.2)]. Note that a valid MATHEMATICA implementation code for evaluating the Fox's H-function has been provided in \cite{26}.

From (4), the E2E SNR when AF relaying scheme is considered can be expressed as [27, Eq. (6)]
\begin{align}
\gamma^{AF}=\frac{\gamma_{1}\gamma_{2}}{\gamma_{2}+C},
\tag{9}\label{9}
\end{align}
where $C$ is a constant that depends on the relay gain, i.e., $C=1/(G^2N_{01})$ [27, Eq. (5)].

By its turn, for the DF relaying scheme, the E2E SNR can be expressed as
\begin{align}
\gamma^{DF}=\min(\gamma_{1},\gamma_{2}).
\tag{10}\label{10}
\end{align}

\section{Closed-Form End-to-End Statistics}
\subsection{Fixed-Gain AF Relaying Protocol}
\subsubsection{Cumulative Distribution Function}
For fixed-gain AF relaying, the CDF of the E2E SNR can be expressed as
\begin{align}
F_{\gamma^{AF}}(\gamma)&=\int_{0}^{\infty}
P\left[\frac{\gamma_{1}\gamma_{2}}{\gamma_{2}+C}
<\gamma \Big|\gamma_{1}\right]f_{\gamma_{1}}(\gamma_{1})d\gamma_{1},
\tag{11}\label{11}
\end{align}
where $F_{\gamma_{1}}(\gamma)$ is the CDF of $\gamma_{1}$. From [28, Eq. (A.3)] and taking the following change of variables $x=\gamma_{1}-\gamma$, (11) can be rewritten as
\begin{align}
F_{\gamma^{AF}}(\gamma)=F_{\gamma_{1}}(\gamma)+\underbrace{\int_{0}^{\infty}F_{\gamma_{2}}
\left(\frac{C\gamma}{x}\right)f_{\gamma_{1}}(x+\gamma)dx}_{\mathcal{I}_{1}}.
\tag{12}\label{12}
\end{align}
By replacing (2) into (12) to calculate $\mathcal{I}_{1}$, a closed-form expression for the CDF of the fixed-gain AF relaying is hard to attain. In order to arrive at a closed-form expression for $\mathcal{I}_{1}$, we rely on [29, Eqs. (13) and (19)] so that (\ref{12}) can be simplified as
\begin{align}
f_{\gamma_{1}}(\gamma_{1})\approx \sum_{m=0}^{n}&a_{1} \gamma_{1}^m\exp\left(-\frac{\vartheta\gamma_{1}}{\overline\gamma_{1}}\right),
\tag{13}\label{13}
\end{align}
where $a_{1}=\varpi \left(\frac{\vartheta}{\overline\gamma_{1}}\right)^{m+1}$ with $\varpi=\frac{(n-m-1)!n^{1-2m}K^{m}e^{-K}}{m!(n-m)!\Gamma(m{+}1)}$. Then, applying [22, Eq. (3.351.1)], the CDF of $\gamma_{1}$ can be approximately derived as
\begin{align}
F_{\gamma_{1}}(\gamma_{1})\approx \sum_{m=0}^{n}&\varpi\Upsilon\left(m+1,\frac{\vartheta \gamma_{1}}{\overline\gamma_{1}}\right),
\tag{14}\label{14}
\end{align}
where $\Upsilon(\cdot,\cdot)$ is the lower incomplete Gamma function [22, Eq. (8.350.1)]. Finally, by substituting (\ref{8}) and (\ref{13}) into $\mathcal{I}_{1}$, and after some algebraic operations, we have
\begin{align}
&\mathcal{I}_{1} =\sum_{m=0}^{n}\epsilon_{1}\gamma^{m+1}{\rm {H}_{1,0:3,1:1,1}^{0,1:1,2:0,1}} \left [{\!\!\left .{ \begin{matrix}
\left({m+2;-1,1}\right)\\ -\\ (0,r)(0,1)(1,r) \\ (0,r)\\ (1,1)
\\ (m+1,1) \end{matrix} }\right |\!
\frac{\lambda^r \mu_{r}}{C}, \!\frac{\overline\gamma_{1}}{\vartheta\gamma} \!\!}\right ]\nonumber\\
&+ \sum_{m=0}^{n}\epsilon_{2}\gamma^{m+1}{\rm {H}_{1,0:3,1:1,1}^{0,1:1,2:0,1}} \left [{\!\!\left .{ \begin{matrix}
\left({m+2;-1,1}\right)\\ -\\ (1{-}a,\frac{r}{c})(0,1)(1,\frac{r}{c}) \\ (0,\frac{r}{c})
\\ (1,1)\\ (m+1,1) \end{matrix} }\right |\!
\frac{b^r \mu_{r}}{C}, \!\frac{\overline\gamma_{1}}{\vartheta\gamma} \!\!}\right ],
\tag{15}\label{15}
\end{align}
where $\epsilon_{1}=a_{1}\omega r$, $\epsilon_{2}=\frac{a_{1}(1-\omega)r}{\Gamma(a)c}$ and
$\rm {H}_{\cdot,\cdot:\cdot,\cdot:\cdot,\cdot}^{\cdot,\cdot:\cdot,\cdot:\cdot,\cdot}[\cdot,\cdot]$
denotes the EGBFHF [25, Eq. (2.56)]. It is worth noting that the EGBFHF can be efficiently calculated
by using the MATLAB implementation code available in \cite{30}.

\textit{Proof:} See Appendix A.

Thus, with the help of (\ref{3}) and (\ref{15}), the CDF of the fixed-gain AF relaying system is finally formulated as
\begin{align}
F_{\gamma^{AF}}(\gamma )=1-Q_{1}\left(\sqrt{2K},\sqrt{2\gamma \vartheta/ \overline\gamma_{1}}\right)+\mathcal{I}_{1}.
\tag{16}\label{16}
\end{align}

\subsubsection{Probability Density Function}
By making the derivative of (\ref{11}), the PDF of the E2E SNR is formulated as
\begin{align}
f_{\gamma^{AF}}(\gamma)=\frac{d}{d\gamma}\int_{0}^{\infty}
P\left[\frac{\gamma_{1}\gamma_{2}}{\gamma_{2}+C}
<\gamma \Big|\gamma_{1}\right]f_{\gamma_{1}}(\gamma_{1})d\gamma_{1}.
\tag{17}\label{17}
\end{align}
Using [19, Eq. (25)] and after some algebraic operations, the final expression for the PDF of $\gamma^{AF}$ can be formulated as
\begin{align}
&f_{\gamma^{AF}}(\gamma)=\sum_{m=0}^{n} \epsilon_{3}\gamma^{m} {\rm {H}_{1,0:2,0:1,1}^{0,1:0,2:0,1}} \left [{\!\!\left .{ \begin{matrix}
\left({m{+}2;-1,1}\right)\\ -\\ (0,r)(1,1) \\ -\\ (1,1) \\ (m{+}2,1)
\end{matrix} }\right |\!\frac{\lambda^r \mu_{r}}{C}, \!
\frac{\overline\gamma_{1}}{\vartheta\gamma} \!\!}\right ]\nonumber\\
&+\sum_{m=0}^{n} \epsilon_{4}\gamma^{m} {\rm {H}_{1,0:2,0:1,1}^{0,1:0,2:0,1}}
\left [{\!\!\left .{ \begin{matrix}\left({m{+}2;-1,1}\right)\\ -
\\ (1{-}a,\frac{r}{c})(1,1) \\ -\\ (1,1) \\ (m{+}2,1) \end{matrix} }\right |\!
\frac{b^r \mu_{r}}{C}, \!\frac{\overline\gamma_{1}}{\vartheta\gamma} \!\!}\right ],
\tag{18}\label{18}
\end{align}
where $\epsilon_{3}=a_{1}\omega$, $\epsilon_{4}=\frac{a_{1}(1-\omega)}{\Gamma(a)}$. Similar derivations can also be found in \cite{31}.

\subsection{DF Relaying Protocol}
\subsubsection{Cumulative Distribution Function}
As for the DF relaying protocol, the CDF of $\gamma^{DF}$ can be written as
\begin{align}
F_{\gamma^{DF}}(\gamma)&=F_{\gamma_{1}}(\gamma)+F_{\gamma_{2}}(\gamma)
-F_{\gamma_{1}}(\gamma)F_{\gamma_{2}}(\gamma),
\tag{19}\label{19}
\end{align}
where $F_{\gamma_{1}}(\gamma)$ and $F_{\gamma_{2}}(\gamma)$ denote
the CDFs of $\gamma_{1}$ and $\gamma_{2}$, respectively. By substituting
(\ref{3}) and (\ref{8}) in (\ref{19}), the CDF of $\gamma^{DF}$ can be formulated in closed-form.
\subsubsection{Probability Density Function}
By taking the derivative of (\ref{19}) with respect to $\gamma$,
the following PDF is obtained
\begin{align}
f_{\gamma^{DF}}(\gamma)=&f_{\gamma_{1}}(\gamma)+f_{\gamma_{2}}(\gamma)
-f_{\gamma_{1}}(\gamma)F_{\gamma_{2}}(\gamma)\nonumber\\
&-f_{\gamma_{2}}(\gamma)F_{\gamma_{1}}(\gamma),
\tag{20}\label{20}
\end{align}
which can be obtained from the PDFs and CDFs of $\gamma_1$ and $\gamma_2$.

\section{Performance Analysis}
In this section, we derive exact closed-form expressions for the OP, ABER, and ACC of the considered system setup assuming fixed-gain AF and DF relays. In addition, in order to clearly see the impact of fading parameters on the overall performance, tight asymptotic expressions at high SNR regime are derived, based on which the respective diversity orders are computed.

\subsection{Fixed-Gain AF Relaying}
\subsubsection{Outage Probability}
The exact closed-form expression for the OP of dual-hop RF-UWOC systems with
fixed-gain AF relaying system is obtained from (\ref{16}), and it is given by
\begin{align}
P_{out}^{AF}=\Pr[\gamma<\gamma_{th}]=F_{\gamma^{AF}}(\gamma_{th}),
\tag{21}\label{21}
\end{align}
where $P_{out}^{AF}$ holds for the CDF of $\gamma^{AF}$ evaluated at the threshold $\gamma_{th}$.

In order to get more useful performance insights, an asymptotic outage expression can be derived as
\begin{align}
P_{out}^{AF,\infty}\approx F_{\gamma_{1}}^{\infty}(\gamma_{th})+\mathcal{I}_{1}^{\infty}(\gamma_{th}),
\tag{22}\label{22}
\end{align}
where $F_{\gamma_{1}}^{\infty}(\cdot)$ is the asymptotic OP of $\gamma_{1}$ and $\mathcal{I}_{1}^{\infty}(\cdot)$ denotes the asymptotic result of $\mathcal{I}_{1}$.
From \cite{32} and [21, Eq. (18)], when $\overline\gamma_{1}\rightarrow \infty$, the $F_{\gamma_{1}}^{\infty}$ can be written as
\begin{align}
F_{\gamma_{1}}^{\infty}\approx \frac{\vartheta e^{-K}\gamma_{th}}{\overline\gamma_{1}}.
\tag{23}\label{23}
\end{align}
Additionally, assuming that $\overline\gamma_{1}\rightarrow \infty$ and employing [25, Eq. (1.2)], $\mathcal{I}_{1}^{\infty}$
can be derived as
\begin{align}
&\mathcal{I}_{1}^{\infty}\underset{\overline\gamma_{1}\gg 1}=\omega r \sum_{m=0}^{n}\varpi \, {\mathrm{H}}_{1,3}^{2,1}
\left [{{\frac{C\vartheta \gamma_{th}}{\lambda^r \mu_{r}\overline\gamma_{1}}}\left |{ \begin{matrix} {(1,r)}
\\ {(1,r)(m+1,1)(0,r)} \\ \end{matrix} }\right . }\right ]\!\nonumber\\
&+\frac{(1-\omega) r}{\Gamma(a)c} \sum_{m=0}^{n}\varpi \, {\mathrm{H}}_{1,3}^{2,1}
\left [{{\frac{C\vartheta \gamma_{th}}{b^r \mu_{r}\overline\gamma_{1}}}\left |{ \begin{matrix} {(1,\frac{r}{c})}
\\ {(a,\frac{r}{c})(m+1,1)(0,\frac{r}{c})} \\ \end{matrix} }\right . }\right ]\!.
\tag{24}\label{24}
\end{align}
When $\overline\gamma_{2}\rightarrow \infty$, $\mathcal{I}_{1}$ can be further asymptotically expressed by using [33, Eq. (1.8.4)]
as
\begin{align}
&\mathcal{I}_{1}^{\infty}\underset{\overline\gamma_{1},~\mu_{r} \gg 1}\approx \omega r \sum_{m=0}^{n}\sum_{j=1}^{2}\varpi \nonumber\\
&\times\frac{\Pi_{i=1,i\neq j}^{2}\Gamma\left(b_{1,i}{-}\frac{b_{1,j}\beta_{1,i}}{\beta_{1,j}}\right)
\Gamma\left(\frac{r b_{1,j}}{\beta_{1,j}}\right)}{\beta_{1,j}\Gamma\left(1{+}\frac{r b_{1,j}}{\beta_{1,j}}\right)} \left(\frac{C\vartheta \gamma_{th}}{\lambda^r \mu_{r}\overline\gamma_{1}}\right)^{\frac{b_{1,j}}{\beta_{1,j}}}\nonumber\\
&+\frac{(1-\omega) r}{\Gamma(a)c} \sum_{m=0}^{n}\sum_{j=1}^{2}\varpi \nonumber\\
&\times \frac{ \Pi_{i=1,i\neq j}^{2}\Gamma\left(b_{2,i}{-}\frac{b_{2,j}\beta_{2,i}}{\beta_{2,j}}\right)
\Gamma\left(\frac{r b_{2,j}}{c\beta_{2,j}}\right)}{\beta_{2,j}\Gamma\left(1{+}\frac{r b_{2,j}}{c\beta_{2,j}}\right)} \left(\frac{C\vartheta \gamma_{th}}{b^r \mu_{r}\overline\gamma_{1}}\right)^{\frac{b_{2,j}}{\beta_{2,j}}},
\tag{25}\label{25}
\end{align}
where $\frac{b_{1,j}}{\beta_{1,j}}=\{m{+}1,\frac{1}{r}\}$ and $\frac{b_{2,j}}{\beta_{2,j}}=\{m{+}1,\frac{ac}{r}\}$. Therefore, by substituting (\ref{23}) and (\ref{25}) into (\ref{22}), the asymptotic OP can be obtained. From Table I and Table II, it is found that $ac>1$. As such, one can conclude that the achievable diversity order of the considered fixed-gain AF relaying system is given by
\begin{align}
G_{d}^{AF} = \min\left(1,\frac{2}{r}\right)=1.
\tag{26}\label{26}
\end{align}
Furthermore, regardless the detection technology used by the R-AUV link, i.e.,  HD (i.e. $r = 1$) or IMDD (i.e. $r = 2$), $G_{d}^{AF}$ is equal to 1. This easy-to-handle result is particularly importance since it reveals that the diversity order of the fixed-gain AF relaying system under study depends solely of the UAV-R link.

\subsubsection{Average Bit Error Rate}
The ABER of various binary modulation schemes can be given by [28, Eq. (25)]
\begin{align}
\overline{P}_{e}=\frac{q^p}{2\Gamma(p)}\int_{0}^{\infty}
\gamma^{p-1}e^{-q\gamma}F_{\gamma}(\gamma)d\gamma,
\tag{27}\label{27}
\end{align}
where the values of $p$ and $q$ determine various modulation scheme.
For instance, $\{p=\frac{1}{2}, q=1\}$ stands for binary phase shift keying (BPSK).
Then, by substituting (\ref{12}) into (\ref{27}), the ABER of fixed-gain AF relaying is formulated as
\begin{align}
\overline{P}_{e}^{AF}{=}&\frac{q^p}{2\Gamma(p)}\int_{0}^{\infty}\gamma^{p{-}1}
e^{-q\gamma}F_{\gamma_{1}}(\gamma)d\gamma\nonumber\\
&+\frac{q^p}{2\Gamma(p)}\int_{0}^{\infty}\gamma^{p{-}1}e^{-q\gamma}\mathcal{I}_{1}d\gamma=\overline{P}_{e,1}+\mathcal{I}_{2},
\tag{28}\label{28}
\end{align}
where $\overline{P}_{e,1}$ is the ABER of the RF link. By replacing (\ref{14}) into (\ref{27}) and utilizing [22, Eq. (6.455.2)], we obtain
\begin{align}
\overline{P}_{e,1}=\sum_{m=0}^{n}&a_{1}\frac{q^p\Gamma(m{+}1{+}p)}{2\Gamma(p)(m{+}1)
\left(q{+}\frac{\theta}{\overline\gamma_{1}}\right)^{m{+}1{+}p}}\nonumber\\
&\times {}_{2}F_{1}\left(1,m{+}1{+}p;m{+}1;\frac{\theta}{q\overline\gamma_{1}{+}\theta}\right).
\tag{29}\label{29}
\end{align}
Based on (\ref{15}), utilizing [22, Eq. (3.326.2)] with the same method in Appendix A, $\mathcal{I}_{2}$ can be formulated as
\begin{align}
&\mathcal{I}_{2}=\sum_{m=0}^{n}\xi_{1} {\rm {H}_{1,0:3,1:1,2}^{0,1:1,2:1,1}} \left [{\!\!\left .{ \begin{matrix}
\left({m{+}2;-1,1}\right)\\ -\\ (0,r)(0,1)(1,r) \\ (0,r)\\ (1,1)\\ (m{+}1{+}p,1)(1,1) \end{matrix} }\right |\!
\frac{\lambda^r \mu_{r}}{C}, \!\frac{q\overline\gamma_{1}}{\vartheta} \!\!}\right ]\nonumber\\
&+\sum_{m=0}^{n}\xi_{2} {\rm {H}_{1,0:3,1:1,2}^{0,1:1,2:1,1}} \left [{\!\!\left .{ \begin{matrix}
\left({m{+}2;-1,1}\right)\\ -\\ (1{-}a,\frac{r}{c})(0,1)(1,\frac{r}{c}) \\ (0,\frac{r}{c})\\ (1,1)\\ (m{+}1{+}p,1)(1,1) \end{matrix} }\right |\!
\frac{b^r \mu_{r}}{C}, \!\frac{q\overline\gamma_{1}}{\vartheta} \!\!}\right ].
\tag{30}\label{30}
\end{align}
where $\xi_{1}=\frac{a_{1}\omega r}{2\Gamma(p)q^{m+1}}$ and $\xi_{2}=\frac{a_{1}(1-\omega) r}{2\Gamma(p)\Gamma(a)cq^{m+1}}$.

When $\overline\gamma_{1}=\overline\gamma_{2}=\overline\gamma\rightarrow \infty$, an asymptotic expression for the ABER is derived as
\begin{align}
\overline{P}_{e}^{AF,\infty}\approx \overline{P}_{e,1}^{\infty}+\mathcal{I}_{2}^{\infty},
\tag{31}\label{31}
\end{align}
where $\overline{P}_{e,1}^{\infty}$ is the asymptotic ABER of the UAV-R link and $\mathcal{I}_{2}^{\infty}$ denotes the asymptotic result of the  $\mathcal{I}_{2}$. By inserting (\ref{23}) into (\ref{27}) and using [22, Eq. (3.326.2)], we obtain
\begin{align}
\overline{P}_{e,1}^{\infty}\approx \frac{\vartheta e^{-K}\Gamma(p+1)}{2\Gamma(p)q\overline\gamma_{1}}.
\tag{32}\label{32}
\end{align}
Moreover, using the same method of $\mathcal{I}_{1}^{\infty}$, $\mathcal{I}_{2}^{\infty}$ can be derived as
\begin{align}
&\mathcal{I}_{2}^{\infty}\underset{\overline\gamma_{1},~\mu_{r} \gg 1}\approx \frac{\omega r}{2\Gamma(p)} \sum_{m=0}^{n}\sum_{j=1}^{2}\varpi \Gamma\left(p{+}\frac{b_{1,j}}{\beta_{1,j}}\right) \nonumber\\
&\times\frac{\Pi_{i=1,i\neq j}^{2}\Gamma\left(b_{1,i}{-}\frac{b_{1,j}\beta_{1,i}}{\beta_{1,j}}\right)
\Gamma\left(\frac{r b_{1,j}}{\beta_{1,j}}\right)}{\beta_{1,j}\Gamma\left(1{+}\frac{r b_{1,j}}{\beta_{1,j}}\right)} \left(\frac{C\vartheta}{\lambda^r q\mu_{r}\overline\gamma_{1}}\right)^{\frac{b_{1,j}}{\beta_{1,j}}}\nonumber\\
&+\frac{(1-\omega) r}{2\Gamma(p)\Gamma(a)c} \sum_{m=0}^{n}\sum_{j=1}^{2}\varpi \Gamma\left(p{+}\frac{b_{2,j}}{\beta_{2,j}}\right)\nonumber\\
&\times \frac{ \Pi_{i=1,i\neq j}^{2}\Gamma\left(b_{2,i}{-}\frac{b_{2,j}\beta_{2,i}}{\beta_{2,j}}\right)
\Gamma\left(\frac{r b_{2,j}}{c\beta_{2,j}}\right)}{\beta_{2,j}\Gamma\left(1{+}\frac{r b_{2,j}}{c\beta_{2,j}}\right)} \left(\frac{C\vartheta}{b^r q \mu_{r}\overline\gamma_{1}}\right)^{\frac{b_{2,j}}{\beta_{2,j}}}.
\tag{33}\label{33}
\end{align}
Therefore, the diversity order calculated through the asymptotic ABER coincide, as expected, with (\ref{26}).
\subsubsection{Average Channel Capacity}
The ACC of the considered system setup operating under both HD and IM/DD can be expressed as [34, Eq. (26)], [35, Eq. (25)]
\begin{align}
\overline{C}=\frac{1}{2\ln(2)}\int_{0}^{\infty}\ln(1+\tau\gamma)f_{\gamma}(\gamma)d\gamma,
\tag{34}\label{34}
\end{align}
where $\tau$ has two potential values such that $\tau=e/(2\pi)$ stands for the IM/DD technique and $\tau=1$ represents the HD technique.

For fixed-gain AF relaying, by inserting (\ref{18}) into (\ref{34}), using [36, Eq. (8.4.6/5)] followed by [25, Eq. (2.9)],
then applying [25, Eq. (2.57)], the ACC can be formulated in closed-form as
\begin{align}
&\overline{C}^{AF}=\sum_{m=0}^{n}\kappa_{1} {\rm {H}_{1,0:2,0:2,2}^{0,1:0,2:1,2}}
\left [{\!\!\left .{ \begin{matrix}
\left({m{+}2;-1,1}\right)\\ -\\ (0,r)(1,1) \\ -
\\ (1,1)(m{+}2,1)\\ (m{+}2,1)(m{+}1,1)
\end{matrix} }\right |\! \frac{\lambda^r \mu_{r}}{C},
\!\frac{\overline\gamma_{1}\tau}{\vartheta} \!\!}\right ]\nonumber\\
&+\sum_{m=0}^{n}\kappa_{2} {\rm {H}_{1,0:2,0:2,2}^{0,1:0,2:1,2}}
\left [{\!\!\left .{ \begin{matrix}
\left({m{+}2;-1,1}\right)\\ -\\ (1{-}a,\frac{r}{c})(1,1)
\\ -\\ (1,1)(m{+}2,1)\\ (m{+}2,1)(m{+}1,1)
\end{matrix} }\right |\! \frac{b^r \mu_{r}}{C},
\!\frac{\overline\gamma_{1}\tau}{\vartheta} \!\!}\right ].
\tag{35}\label{35}
\end{align}
where $\kappa_{1}=\frac{a_{1}\omega }{2\ln(2)\tau^{m+1}}$ and $\kappa_{2}=\frac{a_{1}(1-\omega)}{2\ln(2)\Gamma(a)\tau^{m+1}}$.

\subsection{DF Relaying}
\subsubsection{Outage Probability}
For DF relaying, the OP is obtained from (\ref{19}) as
\begin{align}
P_{out}^{DF}=F_{\gamma^{DF}}(\gamma_{th}).
\tag{36}\label{36}
\end{align}

Then, the asymptotic CDF of $\gamma^{DF}$ can be expressed as
\begin{align}
P_{out}^{DF,\infty} \approx F_{\gamma_{1}}^{\infty}(\gamma_{th})
+F_{\gamma_{2}}^{\infty}(\gamma_{th}),
\tag{37}\label{37}
\end{align}
where $F_{\gamma_{1}}^{\infty}(\gamma_{th})$ is the asymptotic CDF of the UAV-R link,
which is given by (\ref{23}). When $\overline\gamma_{2} \rightarrow \infty$,
$F_{\gamma_{2}}^{\infty}(\gamma_{th})$ can be simplified by using [33, Eq. (1.8.4)], which yields
\begin{align}
F_{\gamma_{2}}^{\infty}(\gamma_{th})=\omega
\left(\frac{\gamma_{th}}{\lambda^r \mu_{r}}\right)^{\frac{1}{r}}
+\frac{(1-\omega)}{\Gamma(a+1)}
\left(\frac{\gamma_{th}}{b^r \mu_{r}}\right)^{\frac{ac}{r}},
\tag{38}\label{38}
\end{align}
Therefore, with the aid of (\ref{23}) and (\ref{38}), the diversity order is
\begin{align}
G_{d}^{DF} = \min\left(1,\frac{1}{r}\right).
\tag{35}\label{35}
\end{align}
One can see that the diversity order of the DF relaying is determined by detection schemes used in the R-AUV link.

\subsubsection{Average Bit Error Rate}
The ABER can be expressed as
\begin{align}
\overline{P}_{e}^{DF}&=\overline{P}_{e,1}+\overline{P}_{e,2}-2\overline{P}_{e,1}\overline{P}_{e,2},
\tag{36}\label{36}
\end{align}
where $\overline{P}_{e,2}$ denotes the ABER of the R-AUV link. In particular, $\overline{P}_{e,1}$ is
given in (\ref{26}). By substituting (\ref{8}) into (\ref{27}) and representing $\exp(-q\gamma)$ in terms
of the Fox's H-function using [36, Eq. (8.4.3/1)] and [37, Eq. (07.34.26.0008.01)], i.e., $\exp(-q\gamma){=}\, {\rm {H}}_{0,1}^{1,0}\left [{q\gamma \left |{ \begin{matrix} {-}
\\ {(0,1)} \\ \end{matrix} }\right . }\right ]\!$, and then utilizing [35, Eq. (2.25.1/1)], we obtain
\begin{align}
\overline{P}_{e,2}&=\frac{\omega r}{2\Gamma(p)}\, {\rm {H}}_{2,2}^{1,2}
\left [{{\frac{1}{\lambda^r\mu_{r}q}}\left |{ \begin{matrix} {(1,r)(1{-}p,1)}
\\ {(1,r)(0,r)} \\ \end{matrix} }\right . }\right ]\! \nonumber\\
&+\frac{(1-\omega)r}{2\Gamma(p)\Gamma(a)c}\, {\rm {H}}_{2,2}^{1,2}
\left [{{\frac{1}{b^r\mu_{r}q}}\left |{ \begin{matrix} {(1,\frac{r}{c})(1{-}p,1)}
\\ {(a,\frac{r}{c})(0,\frac{r}{c})} \\ \end{matrix} }\right . }\right ]\!.
\tag{39}\label{39}
\end{align}
Finally, with the help of (\ref{26}), (\ref{39}) and (\ref{36}), a analytical expression for the
ABER can be derived.

Similarly, the asymptotic ABER expression can be formulated as
\begin{align}
\overline{P}_{e}^{DF,\infty}\approx \overline{P}_{e,1}^{\infty} + \overline{P}_{e,2}^{\infty},
\tag{37}\label{37}
\end{align}
where $\overline{P}_{e,1}^{\infty}$ is given by (\ref{32}). Using [33, Eq. (1.8.4)],
the asymptotic expression of $\overline{P}_{e,2}^{\infty}$ is obtained as
\begin{align}
\overline{P}_{e,2}^{\infty}=\frac{\omega}{2\Gamma(p)}
&\Gamma\left(p+\frac{1}{r}\right)\left(\frac{1}{\lambda^r \mu_{r}q}\right)^{\frac{1}{r}}\nonumber\\
&+\frac{(1-\omega)\Gamma\left(p+\frac{ac}{r}\right)}{2\Gamma(p)\Gamma(a+1)}
\left(\frac{1}{b^r \mu_{r}q}\right)^{\frac{ac}{r}}.
\tag{40}\label{40}
\end{align}
Thus, one can see that the diversity order of the considered DF relay is consistent with (\ref{35}).
\subsubsection{Average Channel Capacity}
Substituting (\ref{20}) into (\ref{34}) and applying the identity $\ln(1+\tau\gamma)=\, {\rm {H}}_{2,2}^{1,2}\left [{{\tau\gamma}\left |
{ \begin{matrix} {(1,1)(1,1)}\\{(1,1)(0,1)} \\ \end{matrix} }\right . }\right ]\!$ [36, Eq. (8.4.6/5)] [37, Eq. (07.34.26.0008.01)], the ACC can be formulated as
\begin{align}
\overline{C}^{DF}=&\frac{1}{2\ln(2)}\int_{0}^{\infty}\, {\rm {H}}_{2,2}^{1,2}
\left [{{\tau\gamma}\left |{ \begin{matrix} {(1,1)(1,1)}\\{(1,1)(0,1)} \\
\end{matrix} }\right . }\right ]\!f_{\gamma}^{DF}(\gamma)d\gamma\nonumber\\
=&I_{C1}+I_{C2}-I_{C3}-I_{C4}.
\tag{41}\label{41}
\end{align}
Then, by applying [36, Eq. (2.25.1/1)] and after some algebraic operations, we have
\begin{align}
I_{C1}&=\frac{\varpi}{2\ln(2)}\nonumber \sum_{m=0}^{n} \, {\rm {H}}_{3,2}^{1,3}\left [{{\frac{\overline\gamma_{1}\tau}{\vartheta}}
\left |{ \begin{matrix} {(-m,1)(1,1)(1,1)}\\{(1,1)(0,1)}
\\ \end{matrix} }\right . }\right ]\!.
\tag{42}\label{42}
\end{align}
Similarly, $I_{C2}$ can be derived as
\begin{align}
&I_{C2}=\frac{\omega r}{2\ln(2)}\, {\rm {H}}_{1,2}^{2,1}
\left [{{\frac{1}{\lambda^r\mu_{r}\tau}}\left |{ \begin{matrix} {(0,1)}
\\ {(0,r)(0,1)} \\ \end{matrix} }\right . }\right ]\!\nonumber\\
&+\frac{(1-\omega) r}{2\ln(2)\Gamma(a)c}\, {\rm {H}}_{2,3}^{3,1}
\left [{{\frac{1}{b^r\mu_{r}\tau}}\left |{ \begin{matrix} {(0,1)(1,\frac{r}{c})}
\\ {(a,\frac{r}{c})(0,\frac{r}{c})(0,1)}
\\ \end{matrix} }\right . }\right ]\!.
\tag{43}\label{43}
\end{align}
In addition, with the help of (\ref{8}) and (\ref{13}), and utilizing the identity $\exp\left(-\frac{\vartheta\gamma_{1}}{\overline\gamma_{1}}\right){=}\, {\rm {H}}_{0,1}^{1,0}
\left [{\frac{\vartheta\gamma_{1}}{\overline\gamma_{1}} \left |{ \begin{matrix} {-} \\ {(0,1)}
\\ \end{matrix} }\right . }\right ]\!$ along with [37, Eqs. (1.1) and (2.3)], $I_{C3}$ can be formulated as
\begin{align}
&I_{C3}{=}\sum_{m=0}^{n}\delta_{1} {\rm {H}_{1,0:2,2:1,2}^{0,1:1,2:1,1}}
\left [{\!\!\left .{ \begin{matrix} (-m;1,1)\\ -\\ (1,1)(1,1)
\\ (1,1)(0,1)\\ (1,r)\\ (1,r)(0,r) \end{matrix} }\right |\!
\frac{\overline\gamma_{1}\tau}{\vartheta},
\!\frac{\overline\gamma_{1}}{\lambda^r\mu_{r}\vartheta} \!\!}\right ]\nonumber\\
&{+}\sum_{m=0}^{n} \delta_{2} {\rm {H}_{1,0:2,2:1,2}^{0,1:1,2:1,1}} \left [{\!\!
\left .{ \begin{matrix} (-m;1,1)\\ -\\ (1,1)(1,1) \\ (1,1)(0,1)\\ (1,\frac{r}{c})\\
(a,\frac{r}{c})(0,\frac{r}{c}) \end{matrix} }\right |\! \frac{\overline\gamma_{1}\tau}{\vartheta}, \!
\frac{\overline\gamma_{1}}{b^r\mu_{r}\vartheta} \!\!}\right ],
\tag{44}\label{44}
\end{align}
where $\delta_{1}=\frac{\varpi\omega r}{2\ln(2)}$ and $\delta_{2}=\frac{\varpi(1-\omega)r}{2\ln(2)\Gamma(a)c}$.
With the help of (\ref{7}), (\ref{14}) and (\ref{41}), $I_{C4}$ can be derived as
\begin{align}
&I_{C4}=\sum_{m=0}^{n}\delta_{3} {\rm {H}_{1,0:2,2:1,2}^{0,1:1,2:1,1}}
\left [{\!\!\left .{ \begin{matrix} (0;r,r)\\ -\\ (1,1)(1,1)
\\ (1,1)(0,1)\\ (1,1)\\ (m+1,1)(0,1)
\end{matrix} }\right |\! \tau \lambda^r\mu_{r}, \!
\frac{\lambda^r\mu_{r}\vartheta} {\overline\gamma_{1}} \!\!}\right ]\nonumber\\
&{+}\sum_{m=0}^{n}\delta_{4}{\rm {H}_{1,0:2,2:1,2}^{0,1:1,2:1,1}}
\left [{\!\!\left .{ \begin{matrix} (1{-}a;\frac{r}{c},\frac{r}{c})
\\ -\\ (1,1)(1,1) \\ (1,1)(0,1)\\ (1,1)\\ (m+1,1)(0,1)
\end{matrix} }\right |\! \tau b^r\mu_{r},
\!\frac{b^r\mu_{r}\vartheta}{\overline\gamma_{1}} \!\!}\right ].
\tag{45}\label{45}
\end{align}
where $\delta_{3}=\frac{\varpi\omega}{2\ln(2)}$ and $\delta_{4}=\frac{\varpi(1-\omega)}{2\ln(2)\Gamma(a)}$.

\section{Optimal Altitude Analysis}
In UAV-based communication networks, although high altitude can increase the coverage, it may cause a large path loss. As such, a trade-off exists between link quality and coverage which required a more in-depth investigation. In this section, we take the DF relaying as an example to analyze the optimal deployment of the UAV with the aim to minimize the OP.

Following (\ref{19}), (\ref{21}) and (\ref{36}), the OP of the DF relaying $P_{out}^{DF}$ can be written by using (\ref{3}) and (\ref{8}) as
\begin{align}
&P_{out}^{DF}=1-\left({1-\omega r\, {\mathrm{H}}_{1,2}^{1,1}
\left [{{\frac{\gamma_{th}}{\lambda^r \mu_{r}}}\left |{ \begin{matrix} {(1,r)}
\\ {(1,r)(0,r)} \\ \end{matrix} }\right . }\right ]\!}\right.\nonumber\\
&\left.{-\frac{(1-\omega)r}{\Gamma(a)c}\, {\mathrm{H}}_{1,2}^{1,1}
\left [{{\frac{\gamma_{th}}{b^r \mu_{r}}}\left |{ \begin{matrix} {(1,\frac{r}{c})}
\\ {(a,\frac{r}{c})(0,\frac{r}{c})} \\ \end{matrix} }\right . }\right ]\!}\right)\nonumber\\
&\times Q_{1}\left(\sqrt{2K(\theta)},\sqrt{2\gamma_{th} (1+K(\theta))d_{UR}^{\alpha(\theta)}/ \overline\gamma_{1}}\right).
\tag{46}\label{46}
\end{align}
Note that the OP depends on the UAV altitude $h_{1}$ and the horizontal distance $r_{1}$ \cite{10}. Thus, (\ref{46}) is formulated as $P_{out}^{DF}=P_{out}^{DF}(r_{1}, h_{1})$. At a given parameter $r_{1}$, the optimal altitude can be obtained as
\begin{align}
\widetilde{h} = \arg \underset{h\in[0,\infty)}{\min}P_{out}^{DF}(r_{1}, h_{1}).
\tag{47}\label{47}
\end{align}

Therefore, to minimize the OP, we take the derivative of (\ref{46}) with respect to $\theta$ as
\begin{align}
\frac{\partial}{\partial \theta}P_{out}^{DF}(r_{1}, h_{1}) = 0,
\tag{48}\label{48}
\end{align}
Then, by substituting (\ref{46}) into (\ref{48}), we get
\begin{align}
\Lambda_{1} \frac{\partial Q(x,y)}{\partial \theta}=0,
\tag{49}\label{49}
\end{align}
where $x=\sqrt{2K(\theta)}$ and $y=\sqrt{2\gamma_{th} (1+K(\theta))d_{UR}^{\alpha(\theta)}/ \overline\gamma_{1}}$ represent
the first and second arguments of the Marcum Q-function in (\ref{46}), respectively. From (49), one can obtain the following expression
\begin{align}
\Lambda_{1} = &\omega r\, {\mathrm{H}}_{1,2}^{1,1}
\left [{{\frac{\gamma_{th}}{\lambda^r \mu_{r}}}\left |{ \begin{matrix} {(1,r)}
\\ {(1,r)(0,r)} \\ \end{matrix} }\right . }\right ]\!\nonumber\\
&+\frac{(1-\omega)r}{\Gamma(a)c}\, {\mathrm{H}}_{1,2}^{1,1}
\left [{{\frac{\gamma_{th}}{b^r \mu_{r}}}\left |{ \begin{matrix} {(1,\frac{r}{c})}
\\ {(a,\frac{r}{c})(0,\frac{r}{c})} \\ \end{matrix} }\right . }\right ]\!-1,
\tag{50}\label{50}
\end{align}
which shows that $\Lambda_{1}$ is independent of $\theta$. In fact, by letting $\frac{\partial Q(x,y)}{\partial\theta}=0$, the optimal solution can be obtained.
Therefore, taking the required derivative, it follows that [5, Eq. (38)]
\begin{align}
&\frac{\partial Q(x,y)}{\partial \theta}=ye^{\frac{x^2+y^2}{2}}\left[{I_{1}(xy)\frac{K'(\theta)}{x}-I_{0}(xy)}\right.\nonumber\\
&\left.{\times\frac{y}{2}\left[\frac{K'(\theta)}{1+K(\theta)}+\alpha'(\theta)
\ln\left(\frac{r_{1}}{\cos\theta}\right)+\alpha(\theta)\tan\theta\right]}\right],
\tag{51}\label{51}
\end{align}
where $I_{1}(\cdot)$ denotes the first-order modified Bessel function.

Therefore, for a given $r_{1}$, the optimal altitude $\widetilde{h}$ can be expressed as
\begin{align}
\widetilde{h}_{1} = r_{1}\tan(\widetilde{\theta}),
\tag{52}\label{52}
\end{align}
where $\widetilde{\theta}$ is obtained from (\ref{49}).

\section{Numerical Results and Discussions}
In this section, Monte Carlo simulation is used to verify the correctness of our analysis. Without loss of generality, we set the threshold  $\gamma_{th}=1.5$ dB and the relaying gain is $C=1.3$. Also, we set the average SNR of the UAV-R link and the R-AUV link to be equal, i.e, $\overline\gamma_{1}=\overline\gamma_{2}=\overline\gamma$. The parameters of the UAV-R link are $A=1$, $a_{1}=-1.5$, $b_{1}=3.5$, $K(0)=5$ dB, and $K(\frac{\pi}{2})=15$ dB. In addition, the parameters setup used in the R-AUV link is listed in Table I and Table II.

\begin{table}[t]
\centering
\caption{The Optimal Elevation Angle $\widetilde{\theta}$[$^\circ$] and Optimal Altitude $\widetilde{h}_{1}$[m] at Given $r_{1}$[m] in The Case of Different Turbulence Conditions.}
{\begin{tabular}{p{1.4cm}p{1.8cm}p{1.8cm}p{1.6cm}}
 \hline
 \hline
$r_{1}$[m] & $\widetilde{\theta}$ [$^\circ$], & $\widetilde{h}_{1}$[m], & $\widetilde{h}_{1}$[m], \\
 & $\sigma_{I}^{2}=0.1484$ & $\sigma_{I}^{2}=0.1484$ & $\sigma_{I}^{2}=1.1273$ \\
 \hline
500 & 70.9 & 1443.9 & 1443.9  \\
600 & 69.0 & 1563.1 & 1563.1  \\
700 & 67.1 & 1657.1 & 1657.1  \\
800 & 65.3 & 1793.3 & 1793.3  \\
900 & 63.6 & 1813.0 & 1813.0  \\
1000 & 62.0 & 1880.7 & 1880.7  \\
1100 & 60.4 & 1936.4 & 1936.4  \\
1200 & 58.8 & 1981.4 & 1981.4  \\
1300 & 57.3 & 2025.0 & 2025.0  \\
1400 & 55.8 & 2060.0 & 2060.0  \\
1500 & 54.3 & 2087.5 & 2087.5  \\
 \hline
 \hline
\end{tabular}}
\end{table}

Fig. 2 depicts the OP versus $h_{1}$ of the considered fixed-gain AF relaying with $\overline\gamma=15$ dB and various values of $r_{1}$. Results show that by increasing $h_{1}$, the OP first decreases and then increases. This is because the positioning of the UAV determines the LoS probability and the increase of the UAV altitude implies a higher LoS contribution. As such, the system has an optimal altitude, enabling to minimize the OP. However, if $h_{1}$ continues to increase, the path loss affects the system performance and reduces the link reliability. In addition, one can notice that reducing $r_{1}$ increases the coverage probability of the UAV and improves the system performance. Additionally, Fig. 3 studies the effect of $h_{1}$ on the obtained OP of the DF relaying, where a three-dimensional (3D) plot shows the relation between $h_{1}$, $r_{1}$, and the resulting OP. In this figuer, we can clear see that different distances imply in different optimal altitudes to guarantee the minimum OP.
Furthermore, for a given $r_{1}$, the optimal elevation angles and optimal UAV altitudes $\widetilde{h}_{1}$ enabling to minimize the OP are presented in Table III, considering different turbulence conditions. As can be seen, $\widetilde{\theta}$ reduces with $r_{1}$ since the link length is more susceptible to the elevation angle at large $r_{1}$. Moreover, one can observe that UWOC link parameters have no effect on the optimal altitude.

\begin{figure}[t]
    \centering
    \includegraphics[width=3.5in]{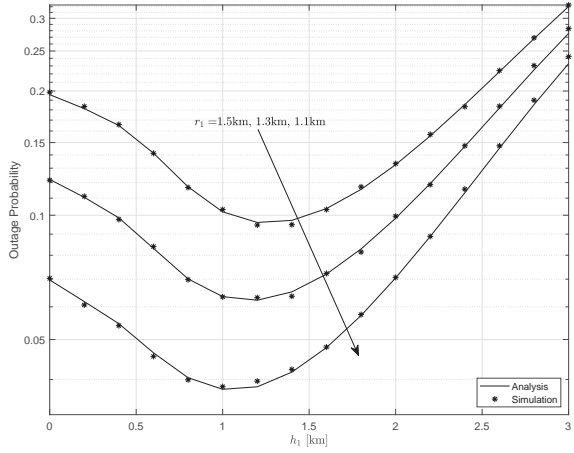}
    \caption{The OP versus $h_{1}$ of the considered fixed-gain AF relaying system with different values of $r_{1}$.}
\end{figure}

\begin{figure}[t]
    \centering
    \includegraphics[width=3.5in]{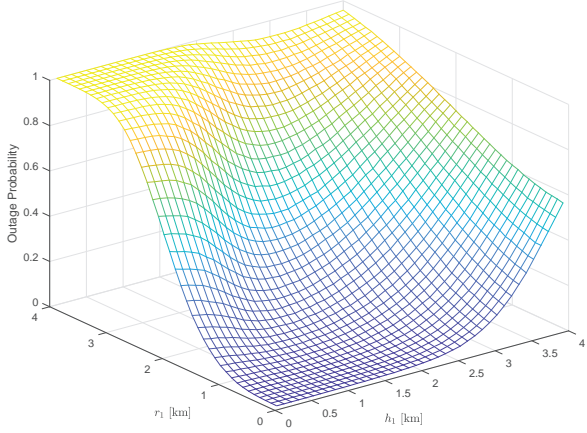}
    \caption{The OP versus $h_{1}$ and $r_{1}$ of the considered fixed-gain AF relaying system.}
\end{figure}

\begin{figure}[t]
    \centering
    \includegraphics[width=3.5in]{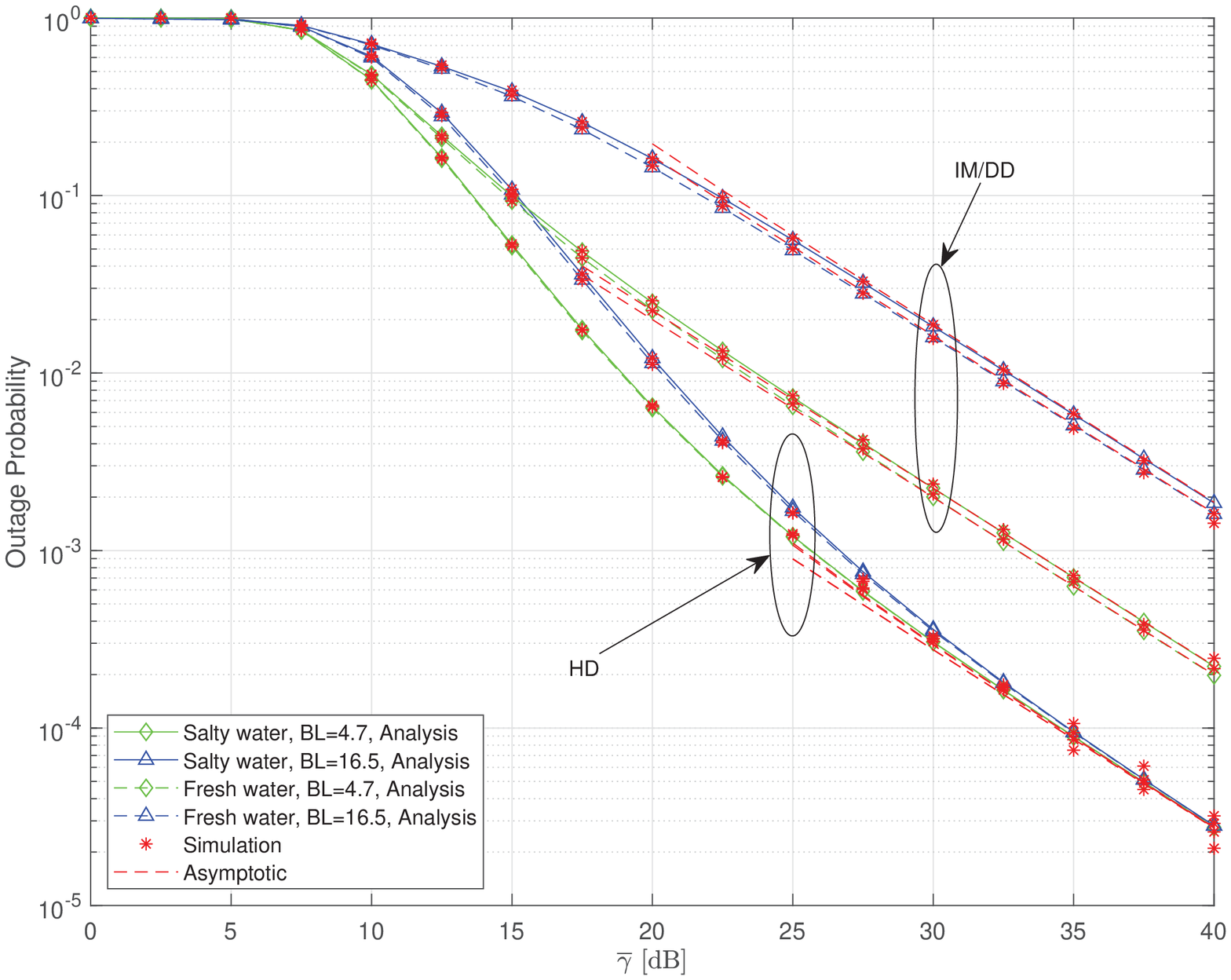}
    \caption{The OP versus $\overline\gamma$ of the fixed-gain AF relaying system for different turbulence conditions and detection techniques, along with the asymptotic results at high SNR.}
\end{figure}

\begin{figure}[t]
    \centering
    \includegraphics[width=3.5in]{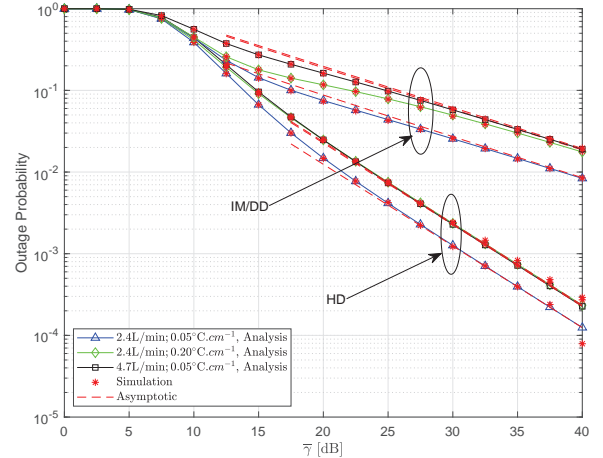}
    \caption{The OP versus $\overline\gamma$ of the DF relaying system for different turbulence conditions and detection techniques, along with the asymptotic results at high SNR.}
\end{figure}

Fig. 4 shows the OP versus $\overline\gamma$ of the considered fixed-gain AF relaying system under different types of water and various levels of air bubbles. In this setup, we assume that $h_{1}=1.5$ km, $r_{1}=1.2$ km. For a fixed type of water, as the bubble level increases, the value of $\sigma_{I}^{2}$ increases, which indicates that the system is experiencing stronger turbulence. In this figure, we assume that BL $=4.7$ L/min corresponds to weak UOT and BL $=16.5$ L/min represents strong UOT. One can observe that for given a type of water, the OP of the system in weak turbulence is significantly lower than that in strong turbulence. If the bubble level is the same, such as BL $=4.7$ L/min, the effect of water salinity on system performance is low. Especially in the case of higher bubble level, i.e, BL $=16.5$ L/min, the influence of salinity is smaller. Therefore, it can be inferred that the impact of bubble level on the considered system is much higher than that of water salinity due to the rapid intensity fluctuations caused by air bubbles. In addition, due to the use of the HD technique can better reduce the performance loss caused by underwater turbulence, note that the outage performance under the HD technique is higher than IM/DD. At high SNR, the asymptotic results are depicted, which reveal that the system diversity order using the two detection techniques is equal. This is because the overall system outage performance is solely determined by the UAV-R link, which is consistent with the result in (\ref{26}).

\begin{figure}[t]
    \centering
    \includegraphics[width=3.5in]{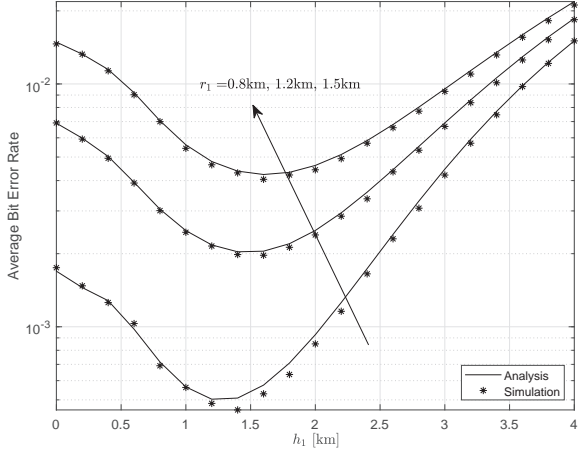}
    \caption{The ABER versus $h_{1}$ of the fixed-gain AF relaying system with different values of the $r_{1}$.}
\end{figure}

\begin{figure}[t]
    \centering
    \includegraphics[width=3.5in]{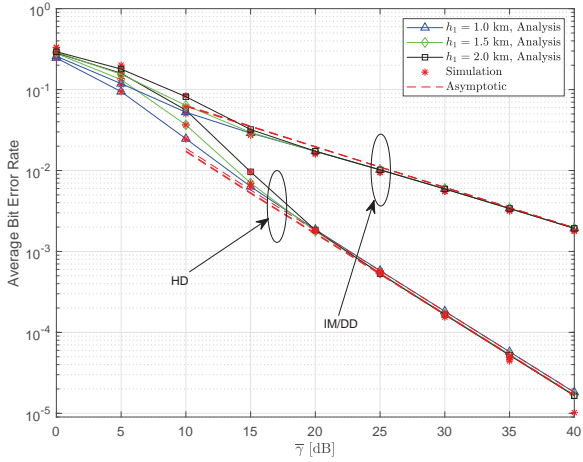}
    \caption{The ABER versus $\overline\gamma$ of the DF relaying system for different altitude of the UAV with $r_{1}=0.8$ km.}
\end{figure}

Fig. 5 shows the OP versus $\overline\gamma$ of the considered DF relaying system under various levels of air bubbles and gradient temperatures for the HD and IM/DD techniques. As can be observed, the HD technique outperforms IM/DD technique, as expected. Moreover, Fig. 5 demonstrates that the bubble levels and gradient temperatures have a serious impact on system outage performance. In particular, the increase in bubble level and gradient temperature causes an increase of the value of $\sigma_{I}^{2}$ and therefore the system suffers from stronger underwater turbulence, which increases the OP. Note also that, under low SNR, air bubbles have a greater adverse effect on outage performance than gradient temperatures. Similarly, the asymptotic results are also shown in this figure, but, different from the fixed-gain AF relaying system, the system diversity order depends on the detection techniques, which is consistent with (\ref{35}).

In Fig. 6, the ABER versus $h_{1}$ is presented for the fixed-gain AF relaying and HD technique by considering different altitude of the UAV. The ABER performance first improves as $h_{1}$ increases and then decreases as the altitude continues to increase. Furthermore, as the $r_{1}$ increases, the error performance gets worse. This is because increasing $r_{1}$ reduces the angle between the UAV and the relay, thereby reducing the probability of the LoS signals.

\begin{figure}[t]
    \centering
    \includegraphics[width=3.5in]{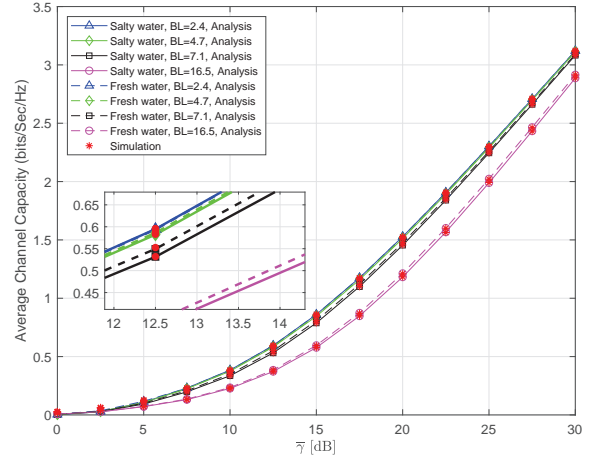}
    \caption{The ACC versus $\overline\gamma$ of the fixed-gain AF relaying system for different levels of air bubbles under IM/DD techniques.}
\end{figure}

\begin{figure}[t]
    \centering
    \includegraphics[width=3.5in]{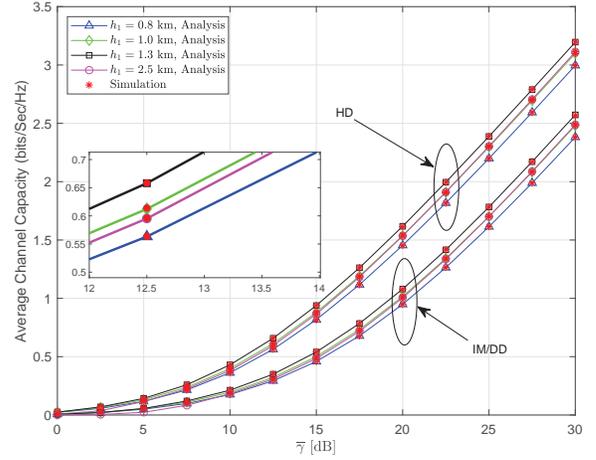}
    \caption{The ACC versus $\overline\gamma$ of the fixed-gain AF relaying system for different altitude of the UAV with $r_{1}=2$ km.}
\end{figure}

In Fig. 7, the ABER versus $\overline\gamma$ is plotted along with the asymptotic results for DF relaying by considering various values of $h_{1}$ and by setting $r_{1}=0.8$ km. Moreover, we assume that the system employs both IM/DD and HD techniques, and it is considered a weak turbulence corresponding to BL $=2.4$ L/min and a temperature gradient of $0.05 ^{\circ}\mathrm{C}.cm^{-1}$. As can be clearly seen, $h_{1}$ can effect the ABER performance at low SNRs. For $r_{1}=0.8$ km, as the UAV altitude drops from 2 km to 1 km, the path loss is reduced, which improves error performance. Furthermore, one can observe that a system with small $h_{1}$ has the same ABER at high SNRs. This is because with the increase of the average SNR and decrease of $h_{1}$, the R-AUV link becomes dominant at high SNR regions.

Assuming various levels of air bubbles and employing salty water and fresh water, Fig. 8 shows the ACC versus $\overline\gamma$ of the fixed-gain AF relaying system in the case of IM/DD technique. As can be clearly seen, as the air bubbles level increases, the ACC decreases. As expected, the performance in slaty water is not as good as that in fresh water, but compared with bubble levels, the water salinity has little effect on performance, especially in the case of weak turbulence, such as BL $=2.4$ L/min.

In Fig. 9, the ACC of the fixed-gain AF relaying is presented versus $\overline\gamma$ for various altitude of the UAV and assuming both IM/DD and HD techniques. In this setup, we set $r_{1}=2$ km, the turbulence condition corresponds to a BL $=2.4$ L/min, and the temperature gradient is $0.05 ^{\circ}\mathrm{C}.cm^{-1}$. As expected, the ACC decreases as $h_{1}$ increases. This is because that increasing the height of the UAV increases the LoS propagation probability of the signal and then improves the ACC. Furthermore, as $h_{1}$ continues to increase, the ACC reduces due to the serious path loss, which is also reflected in Figs. 2 and 6.

\section{Conclusions}
In this paper, we have investigated the performance of the UAV-based dual-hop RF-UWOC system employing fixed-gain AF relaying and DF relaying.
In the UWOC link, both HD and IM/DD modulation schemes were considered. We derived closed-form expressions of the for the OP, ABER, and ACC. For DF relaying, analytical expressions for optimal elevation angle and optimal height were derived. Results showed that UAV can be flexibly deployed as a base station, which brings better performance for dual-hop RF-UWOC systems. It was found that different horizonal distances have different optimal altitudes that guarantee better performance. In addition, temperature fluctuations, salinity variations, and the presence of bubbles all affect the system performance. Specifically, the increase of bubbles levels may cause significant deterioration of the performance compared to changes in temperature and salinity. Furthermore, asymptotic results for the OP and ABER of fixed-gain AF and DF relays were derived and some useful insights were obtained. It was demonstrated that the diversity order of the fixed-gain AF relaying is determined by the RF link, while the diversity order of the DF relaying is determined by the type of detection technique being used in UWOC link.

\appendices
\section{CDF of The E2E SNR for Fixed-Gain AF Relaying}
Making use of [25, Eq. (1.2)] and then [23, Eqs. (3.194.3) and (8.384.1)], we obtain
\begin{align}
&\mathcal{I}_{1} =\sum_{m=0}^{n}a_{1}\omega r\frac{\gamma^{m+1}}{(2\pi i)^2}{\int\limits_{\ell_{1}}}{\int\limits_{\ell_{2}}}
\Gamma(t{-}s{-}m{-}1)\nonumber\\
&\times \frac{\Gamma(1{+}rs)\Gamma(-rs)\Gamma(1{+}s)}{\Gamma(1{-}rs)}\frac{\Gamma(t)}{\Gamma(t{-}m)}\left(\frac{\lambda^r \mu_{r}}{C}\right)^{s} \left(\frac{\overline\gamma_{1}}{\vartheta \gamma}\right)^{t}dsdt\nonumber\\
&+ \sum_{m=0}^{n}a_{1}\frac{(1-\omega) r}{\Gamma(a)c}\frac{\gamma^{m+1}}{(2\pi i)^2}{\int\limits_{\ell_{1}}}{\int\limits_{\ell_{2}}}
\Gamma(t{-}s{-}m{-}1)\nonumber\\
&\times\frac{\Gamma(a{+}\frac{r}{c}s)\Gamma(-\frac{r}{c}s)\Gamma(1{+}s)}{\Gamma(1{-}\frac{r}{c}s)}\frac{\Gamma(t)}{\Gamma(t{-}m)}\left(\frac{b^r \mu_{r}}{C}\right)^{s} \left(\frac{\overline\gamma_{1}}{\vartheta \gamma}\right)^{t}dsdt,
\tag{53}\label{53}
\end{align}
where $\ell_{1}$ and $\ell_{2}$ are the $s-$plane and the $t-$plane contours in
the complex domain, respectively. By utilizing [25, Eqs. (2.56) and (2.57)],
(\ref{53}) can be expressed in closed-form as (\ref{15}).

\end{document}